# Energy Band and Thermodynamics

B.V. Karpenko*, A.V. Kuznetzov** and V.V. Dyakin*

*Institute of physics of metals, Ural department of Russian Academy of Sciences, Ekaterinburg 620041, Russia
**Ural state university, Ekaterinburg 620083, Russia

Address for correspondence: Karpenko Boris Victorovich, Institute of physics of metals, URO RAN, Ekaterinburg 620041, Russia, phone 378-35-64, FACS (343)3745244, e-mail: boris.karpenko@mail.ru

The behavior of various thermodynamic functions in the dependence of degree of energy band occupation and temperature in the one-band tight binding approximation for the crystal was studied. The Fermi energy, density of states, degeneracy temperature, chemical potential, partition function, thermodynamic potential, energy, free energy, entropy, heat capacity, spin magnetization and initial susceptibility were calculated. The limited energetic spectrum leads to the peculiarities in the behavior of these quantities in comparison with free electron gas.

**Аннотация.**
В однозонном приближении сильной связи в кристалле рассмотрено поведение различных термодинамических функций в зависимости от степени заполнения энергетической полосы и температуры. Дан анализ и произведены численные расчеты для энергии Ферми, температуры вырождения, химического потенциала, большой статистической суммы, термодинамического потенциала, энергии, свободной энергии, энтропии, теплоемкости, спиновой намагниченности и начальной восприимчивости. Ограниченность энергетического спектра приводит к особенностям в поведении этих величин по сравнению со свободным электронным газом.

**1. Введение.**

При рассмотрении термодинамики системы делокализованных электронов твердом теле следует описывать их энергетический спектр в терминах зонной теории и применять Фермиевскую статистику. Одним из распространенных подходов является приближение сильной связи. Численный расчет упомянутых термодинамических функций имеет практическое и академическое значение.

**2. Спектр. Концентрация. Функция распределения.**

Известно, что энергия электрона $E$ в приближении сильной связи и при учете соседей первого порядка в простой кубической решетке имеет вид

$$E = \Delta \cdot \varepsilon, \qquad (1)$$

где $\Delta$ есть ширина полосы,

$$\varepsilon = \frac{1}{6}(3 - \cos x - \cos y - \cos z); \quad -\pi \leq x, y, z \leq \pi. \tag{2}$$

($x, y, z$ суть безразмерные квазиимпульсы)

Рассматриваемая модель есть задача с ограниченным спектром.

Ниже мы будем рассматривать систему с произвольной степенью заполнения энергетической полосы, т.е. с электронной концентрацией $n$ при

$$0 \leq n \leq 2 \tag{3}$$

на один узел решетки с учетом спина. Фермиевскую функцию распределения запишем в приведенных переменных:

$$f = \frac{1}{1 + \exp\left(\frac{1}{t}\right)(\varepsilon - w)}, \quad t = \frac{kT}{\Delta}, \quad w = \frac{\mu}{\Delta}, \tag{4}$$

где $\mu$ есть химический потенциал, а $\varepsilon$ дается в (2). Связь между $w$ и $n$ дается обычным соотношением

$$\frac{2}{(2\pi)^3} \int_{-\pi}^{\pi} \int_{-\pi}^{\pi} \int_{-\pi}^{\pi} f\, dx\, dy\, dz = n. \tag{5}$$

### 3. Энергия Ферми, плотность состояний.

Энергия Ферми $\varepsilon_F$ определяется уравнением

$$n = \frac{2}{(2\pi)^3} \iiint dx\, dy\, dz, \tag{6}$$

где интегралы берутся по объему, заключенному изоэнергетической поверхностью $\varepsilon_F$. Функция $\varepsilon_F(n)$ представлена на Рис. 1.

Плотность состояний определяем по формуле

$$D(\varepsilon) = \frac{1}{8\pi^3} \int \frac{dS}{|\nabla \varepsilon|}. \tag{7}$$

В (7) интеграл берется по изоэнергетической поверхности $\varepsilon$. График $D(\varepsilon)$ на Рис. 2.

### 4. Термодинамические функции.

Температура вырождения $t_g$ определяется из условия равенства нулю химического потенциала $w$. На Рис. 3 показана зависимость температуры вырождения от концентрации.

Химический потенциал $w$ определяется из уравнения (5). Семейство кривых $w(t,n)$ изображено на Рис. 4. При $n=2$ химпотенциал есть расходящаяся величина.

Термодинамический потенциал $\Omega$ определяется выражением

$$\Omega = -2\frac{t}{(2\pi)^3} \int_{-\pi}^{\pi}\int_{-\pi}^{\pi}\int_{-\pi}^{\pi} \ln\left(1+\exp\frac{1}{t}(w-\varepsilon)\right) dxdydz. \qquad (8)$$

Кривые $\Omega(t,n)$ показаны на Рис. 5. $\Omega$ расходится при $n=2$.

Для большой статистической суммы $Z$ имеем

$$\ln Z = -\frac{1}{t}\Omega. \qquad (9)$$

Ее график на Рис. 6. И вновь расходимость при $n=2$.

Энергию $E$ находим из формулы

$$E = \frac{2}{(2\pi)^3} \int_{-\pi}^{\pi}\int_{-\pi}^{\pi}\int_{-\pi}^{\pi} \varepsilon f dxdydz. \qquad (10)$$

Графики $E(n,t)$ приведены на Рис. 7.

Свободная энергия $F$ определяется как

$$F = \Omega + wn. \qquad (11)$$

Для удобства графики представлены отдельно для $n \leq 1$ и $n \geq 1$ на Рис. 8 и 9. как функции $t$.

Энтропию $S$ определим по формуле

$$S = \frac{1}{t}(E-F). \qquad (12)$$

На Рис. 10 даны графики $S(t,n)$ для концентраций, дополняющих друг друга до 2.

Теплоемкость $C$ определяется обычным образом:

$$C = \frac{dE}{dt}. \qquad (13)$$

Величина $C(t)$ дается семейством кривых для $n$, дополняющих друг друга до 2, на Рис. 11.

Определим намагниченность $m$ во внешнем магнитном поле $H$ как разность концентраций электронов для спинов, направленных по полю $n_1$ и против него $n_2$:

$$m = n_1 - n_2, \quad (14)$$

$$n_{1,2} = \frac{1}{(2\pi)^3} \int\limits_{-\pi}^{\pi} \int\limits_{-\pi}^{\pi} \int\limits_{-\pi}^{\pi} f_{1,2} dx dy dz, \quad (15)$$

$$f_{1,2} = \frac{1}{1+\exp\frac{1}{t}(\varepsilon_{1,2}-w)}, \quad \varepsilon_1 = \varepsilon - h, \quad \varepsilon_2 = \varepsilon + h, \quad h \equiv \mu_B H, \quad (16)$$

$$n_1 + n_2 = n. \quad (17)$$

При абсолютном нуле температуры зависимости $m(h)$ показаны на Рис. 12. Выход на номинальное значение намагниченности $n$ происходит при полях $h_c$, определяемых формулами: $h_c(n) = \frac{1}{2}\varepsilon_F(2n)$ при $n \leq 1$ и $h_c(n) = \frac{1}{2}\varepsilon_F(2(2-n))$ при $n \geq 1$. Зависимость $h_c(n)$ показана на Рис. 13. При конечных температурах кривые $m(h)$ сглаживаются.

Начальная восприимчивость $\chi_0$ определяется как

$$\chi_0 = \frac{dm}{dh} \text{ при } h=0. \quad (18)$$

Вычисления дают

$$\chi_0 = \frac{K}{t}, \quad (19)$$

где

$$K = \frac{2}{(2\pi)^3} \int\limits_{-\pi}^{\pi} \int\limits_{-\pi}^{\pi} \int\limits_{-\pi}^{\pi} \frac{\exp\left(\frac{1}{t}(\varepsilon-w)\right)}{\left(1+\exp\left(\frac{1}{t}\right)(\varepsilon-w)\right)^2}. \quad (20)$$

При получении этого выражения мы использовали равенство

$$\frac{dw}{dh} = 0 \text{ при } h=0. \quad (21)$$

В (20) химпотенциал берется в нулевом поле. Функции $\chi_0(n,t)$ представлены на Рис. 14. На Рис. 15 те же функции приводятся при малых $t$. Для сравнения приведем график $\chi_0(t)$ для свободного электронного газа на Рис. 16.

### 5. Обсуждение.

Термодинамика системы с ограниченным энергетическим спектром имеет ряд особенностей по сравнению с системами без этого ограничения. Однако эти особенности имеют прозрачный физический смысл и часто предсказуемы.

Концентрация $n=1$ является в некотором смысле "критической". При этой концентрации происходит расходимость температуры вырождения $t_g$ (см. Рис.3). При ней же химпотенциал $w$ не зависит от температуры, оставаясь все время равным 0.5 (см. Рис.4). Химпотенциал может либо падать с температурой ($n<1$), либо расти ($n>1$). Для свободного электронного газа он только падает.

Для концентраций, симметричных относительно $n=1$, величины $S$, $C$, $m$, $K$ также симметричны, демонстрируя эквивалентность дырок и электронов. Энергия $E$ и энтропия $S$ в отличие от свободного газа насыщаются при росте температуры, а не возрастают неограниченно (см. Рис. 7 и 10). Теплоемкость $C$ имеет максимум при изменении температуры и стремится к нулю при неограниченном ее возрастании (см. Рис. 11) в отличие от свободного газа, где $C$ есть монотонная функция температуры, стремящаяся к насыщению. Начальная восприимчивость имеет максимумы при низких температурах при некоторых значениях $n$, что объясняется своеобразной (по сравнению со свободным газом) зависимостью плотности состояний от энергии (см. Рис. 2, 14, 15, 16). Заметим, что химический потенциал играет особую роль в термодинамике системы, поскольку все другие термодинамические функции выражаются через него. Четыре параметра- ширина полосы, концентрация, температура и магнитное поле – определяют искомые функции.

**Acknowledgements**

The author is indebted to Department of Physical Science and Presidium of Ural Branch of the Russian Academy of Sciences, Program "The Physics of new materials and structures", the Russian Foundation for Basic Research (project N 07 02 000 68).

**Подписи к рисункам.**

Рис.1. Зависимость энергии Ферми $\varepsilon_F$ от концентрации $n$.

Рис.2. График плотности состояний $D(\varepsilon)$.

Рис.3. Температура вырождения $t_g$ как функция концентрации $n$.

Рис.4. Зависимости химпотенциала $w$ от температуры $t$ при различных концентрациях $n$. Кривые располагаются снизу вверх при возрастании концентрации $n = 0.1, 0.2..1.9$ и $1.99$.

Рис.5. Зависимости потенциала $\Omega$ от температуры $t$ при различных концентрациях $n$. Кривые располагаются сверху вниз при возрастании концентрации $n = 0.1, 0.2...1.9$ и $1.99$.

Рис.6. Зависимость $\ln Z$ от температуры $t$ при различных концентрациях $n$. Кривые располагаются снизу вверх при возрастании концентрации $n = 0.1, 0.2...1.9$ и $1.99$.

Рис.7. Зависимость энергии $E$ от температуры $t$ при различных $n$. Кривые располагаются сверху вниз при возрастании концентрации $n = 0.1, 0.2...2$.

Рис.8. Свободная энергия $F$ как функция температуры $t$ при $n \leq 1$. Концентрация возрастает снизу вверх при $t = 0$.

Рис.9. Свободная энергия $F$ как функция температуры $t$ при $n \geq 1$. Кривые располагаются снизу вверх при возрастании концентрации $n = 1, 1.1...2$.

Рис.10. Энтропия $S$ как функция температуры $t$ для $n = 0.1(1.9), 0.2(1.8)...1$ снизу вверх.

Рис.11. Теплоемкость $C(t,n)$ для $n = 0.1(1.9), 0.2(1.8)...1$ снизу вверх.

Рис.12. Зависимости $m(h)$ для $n = 0.1(1.9), 0,2(1.8)...1$.

Рис.13. Зависимость $h_c(n)$.

Рис.14. Зависимости $\chi_0(n,t)$.

Рис.15. Зависимости $\chi_0(n,t)$ при малых $t$.

Рис.16. Зависимость $\chi_0(t)$ для свободного электронного газа.

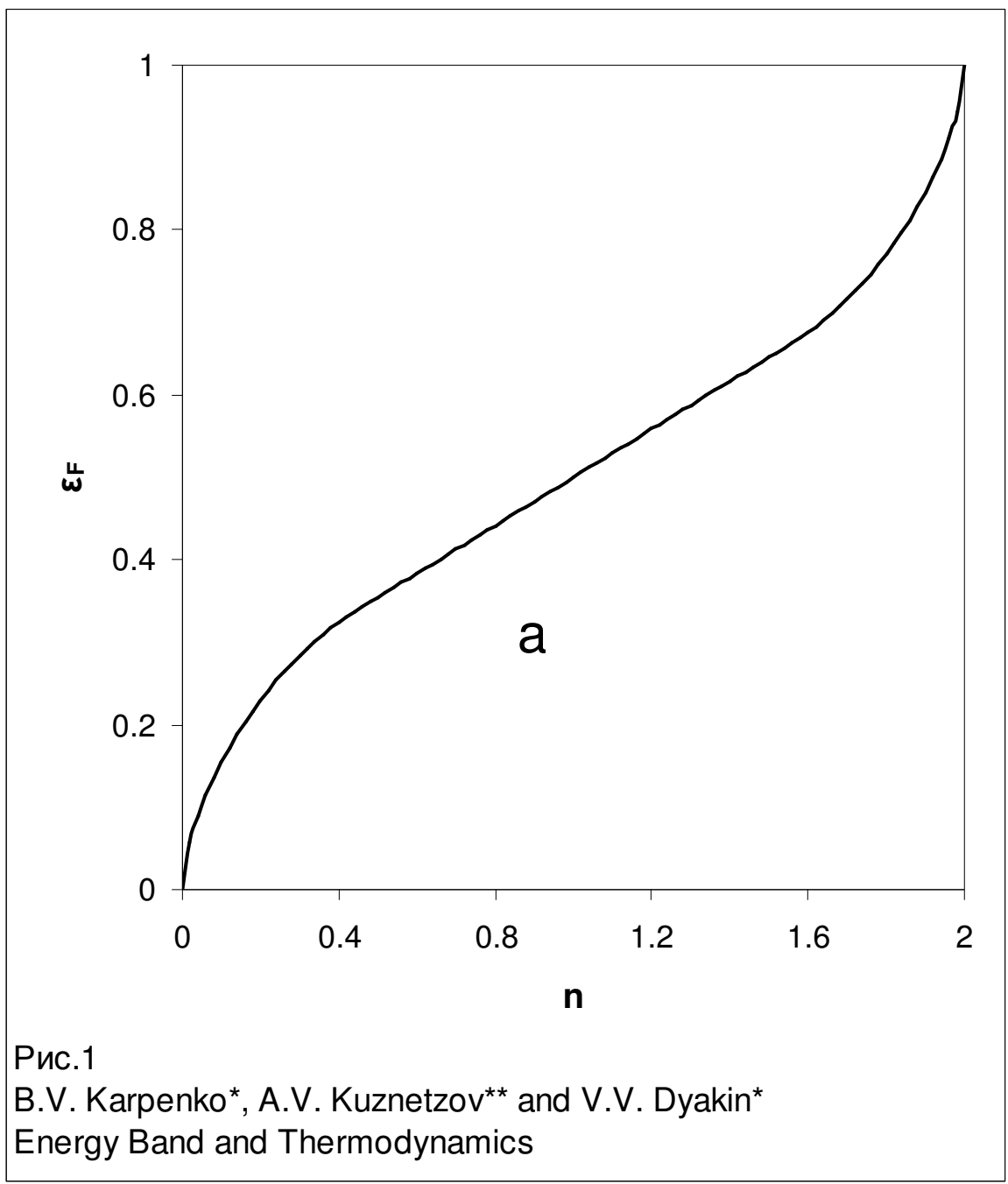

Рис.1
B.V. Karpenko*, A.V. Kuznetzov** and V.V. Dyakin*
Energy Band and Thermodynamics

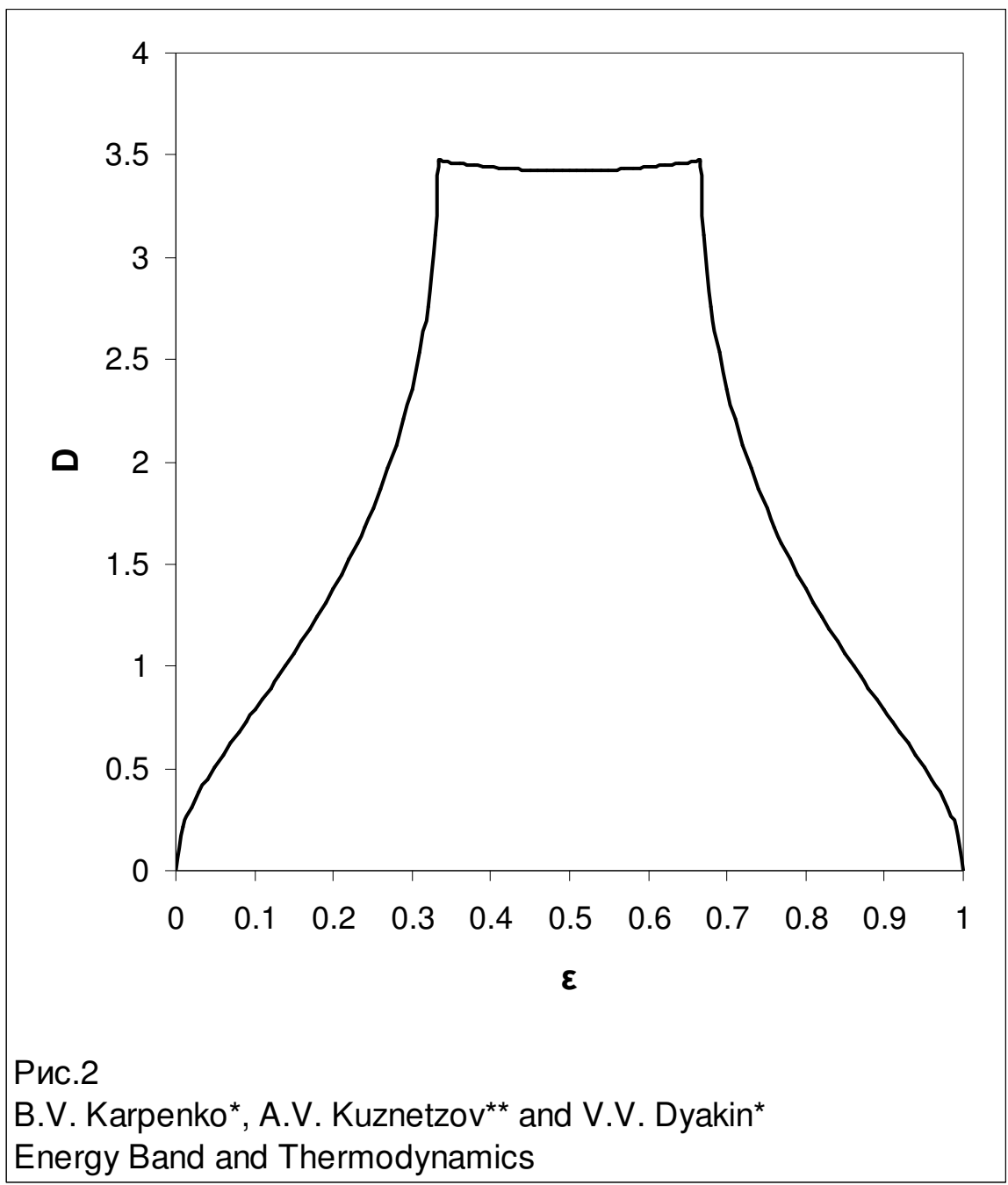

Рис.2
B.V. Karpenko*, A.V. Kuznetzov** and V.V. Dyakin*
Energy Band and Thermodynamics

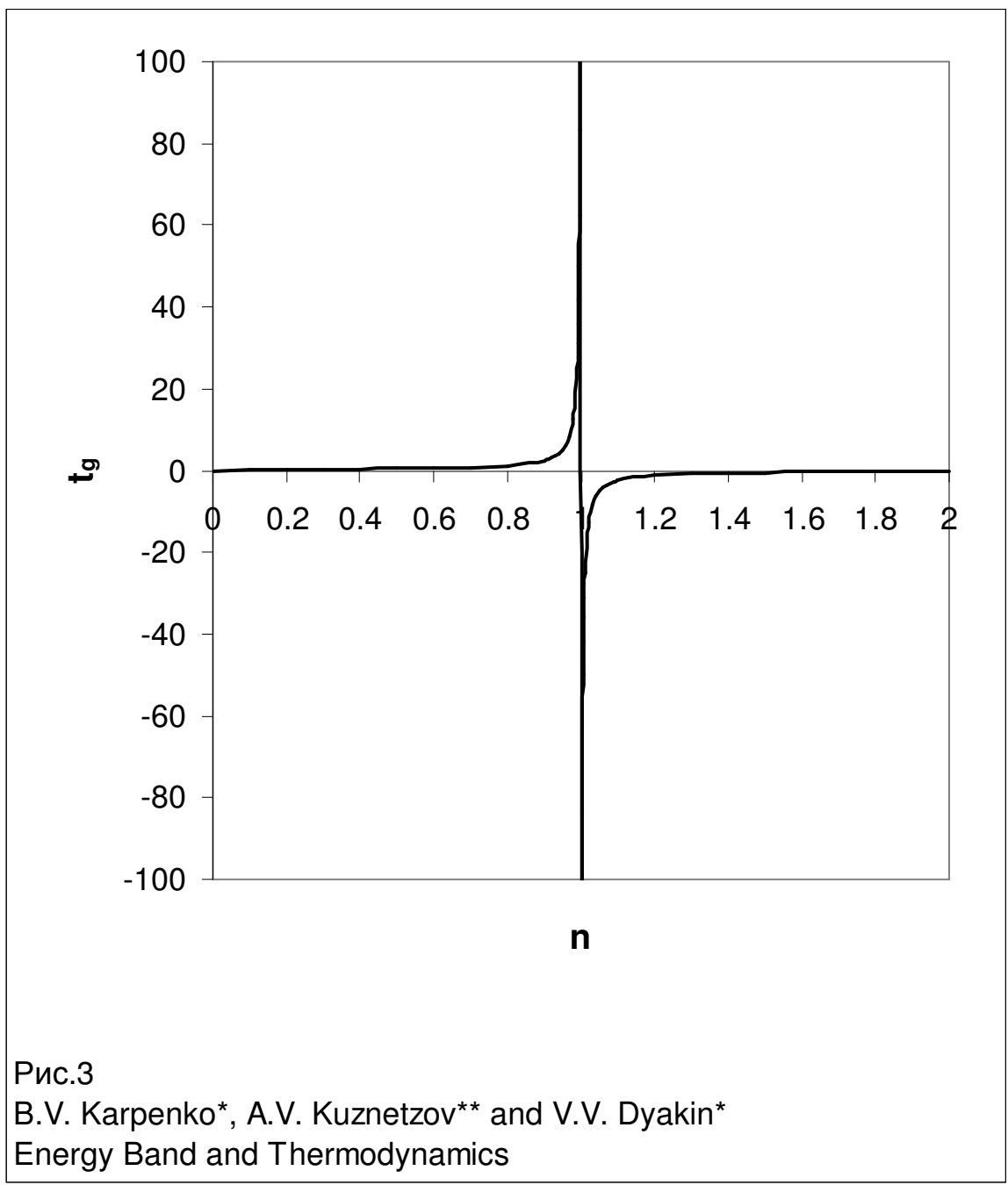

Рис.3
B.V. Karpenko*, A.V. Kuznetzov** and V.V. Dyakin*
Energy Band and Thermodynamics

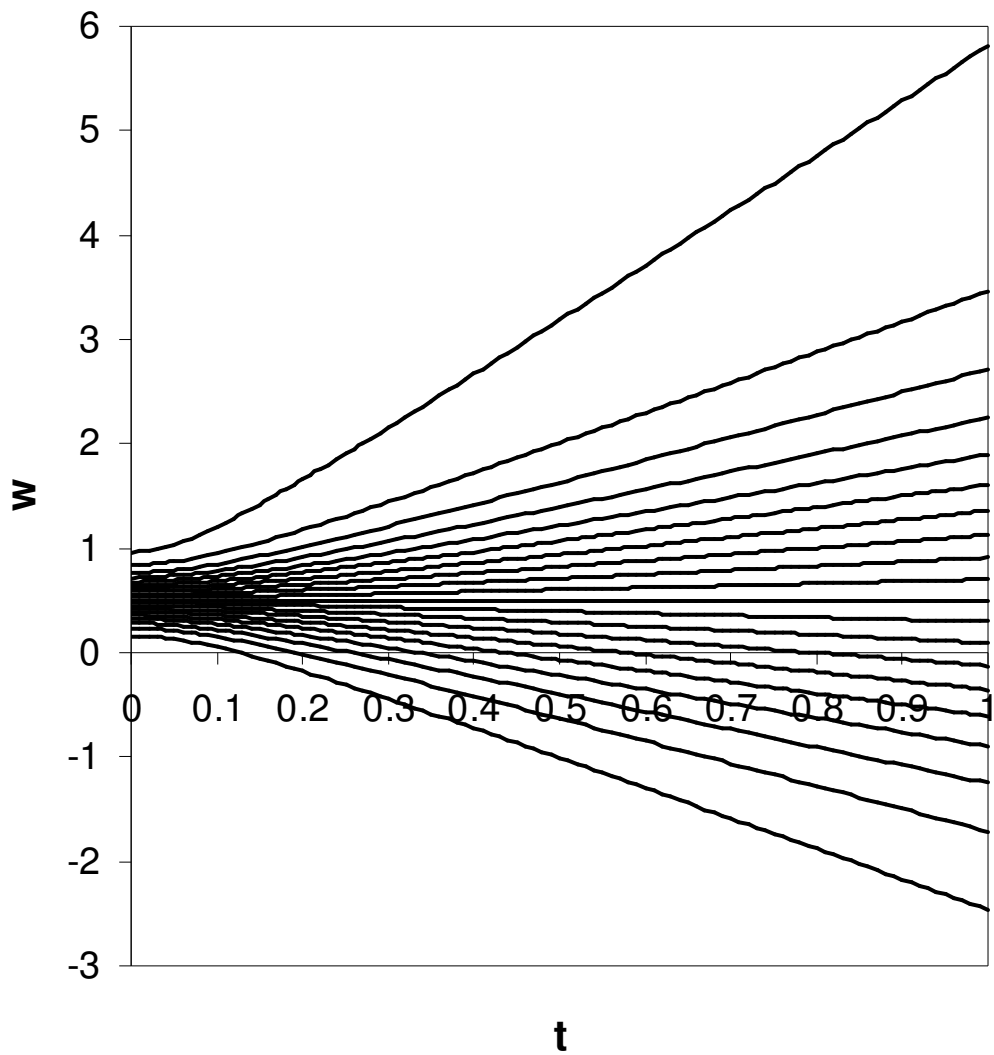

Рис.4
B.V. Karpenko*, A.V. Kuznetzov** and V.V. Dyakin*
Energy Band and Thermodynamics

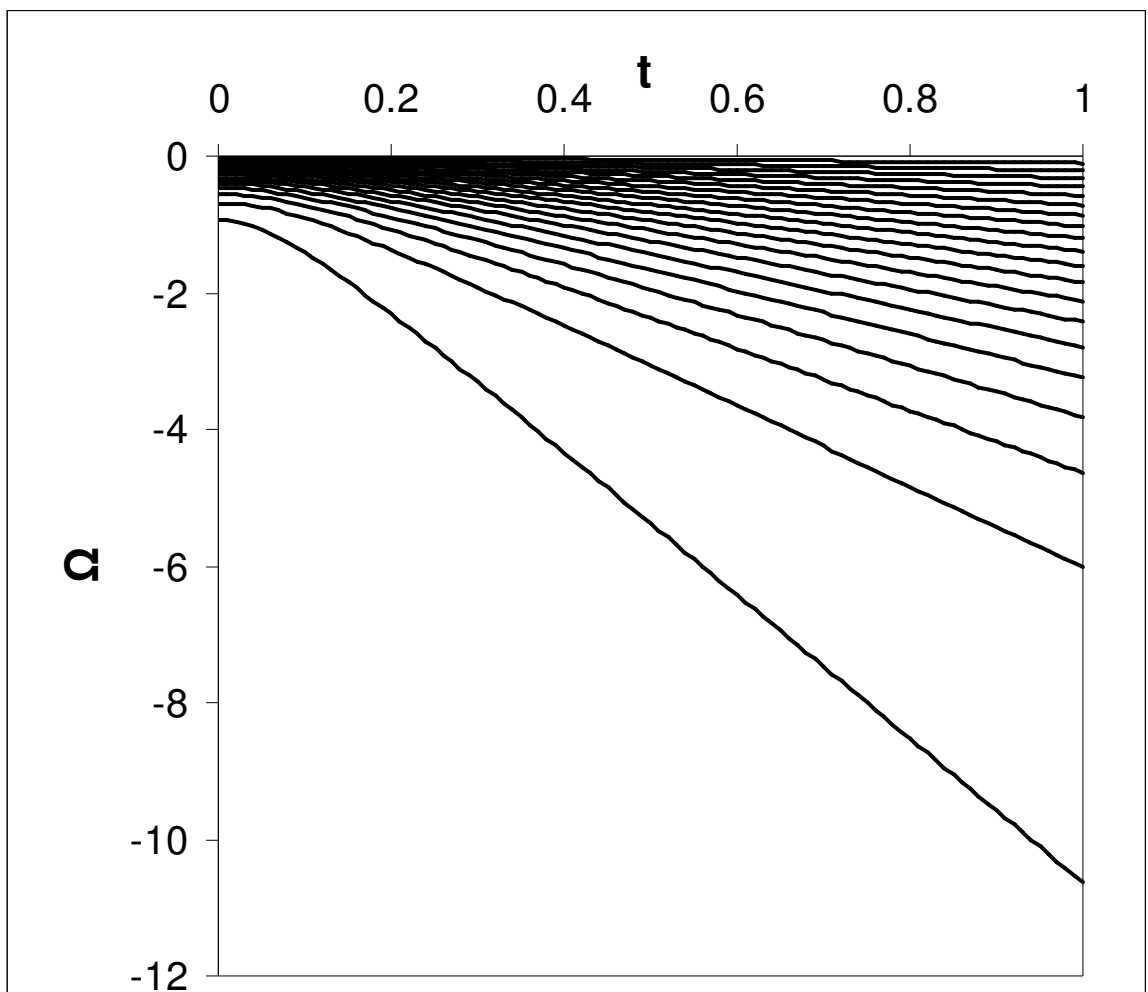

Рис.5
B.V. Karpenko*, A.V. Kuznetzov** and V.V. Dyakin*
Energy Band and Thermodynamics

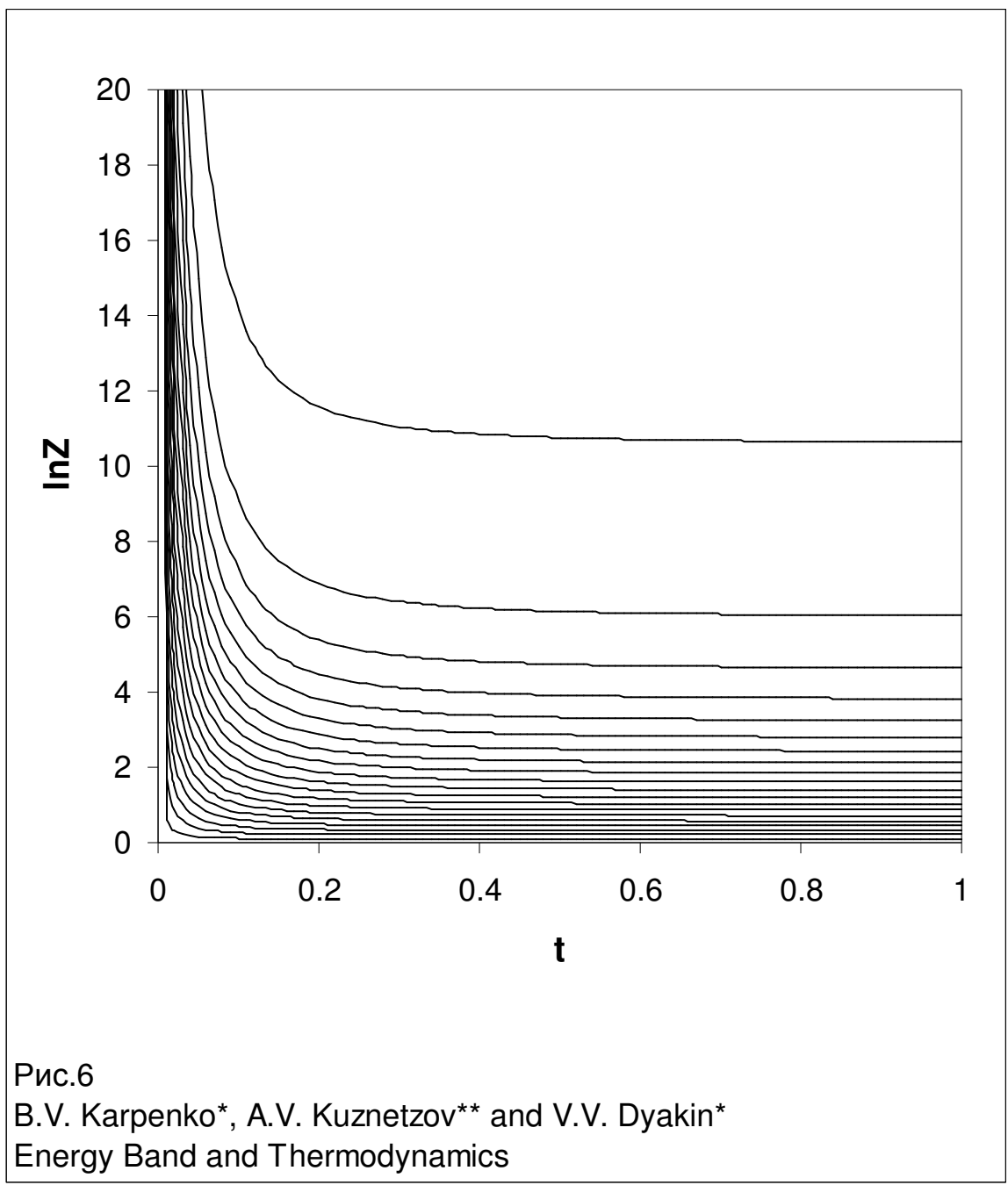

Рис.6
B.V. Karpenko*, A.V. Kuznetzov** and V.V. Dyakin*
Energy Band and Thermodynamics

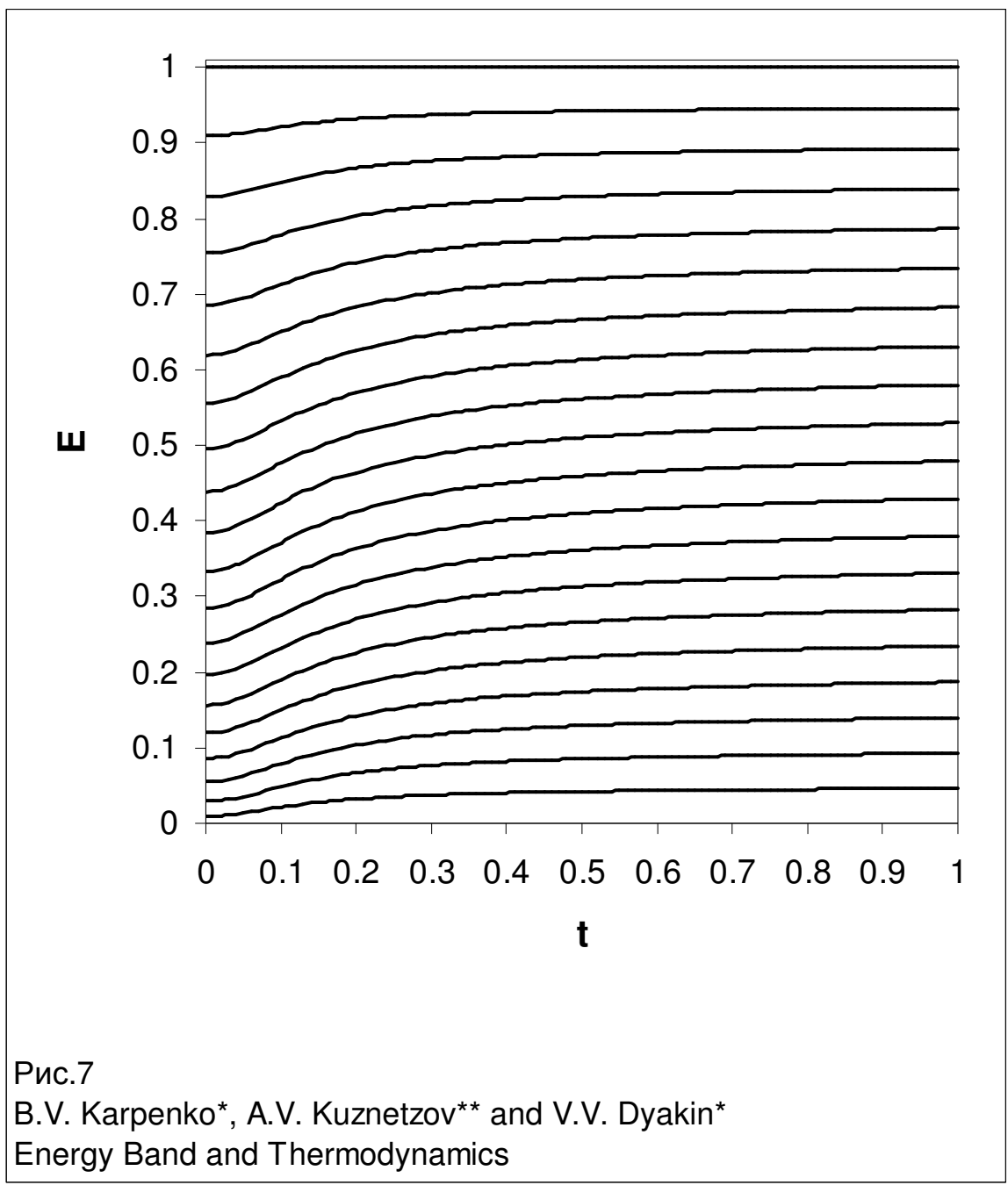

Рис.7
B.V. Karpenko*, A.V. Kuznetzov** and V.V. Dyakin*
Energy Band and Thermodynamics

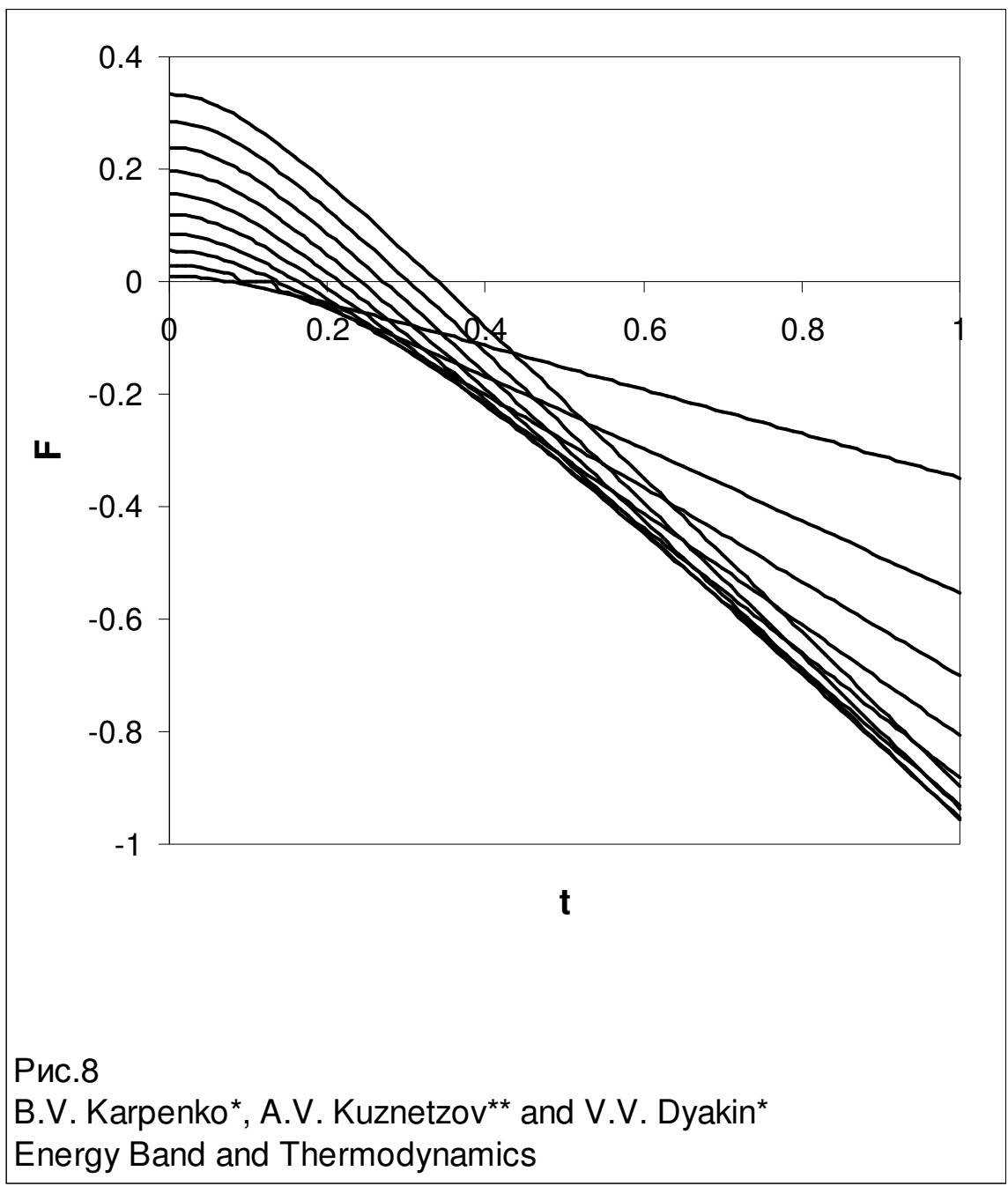

Рис.8
B.V. Karpenko*, A.V. Kuznetzov** and V.V. Dyakin*
Energy Band and Thermodynamics

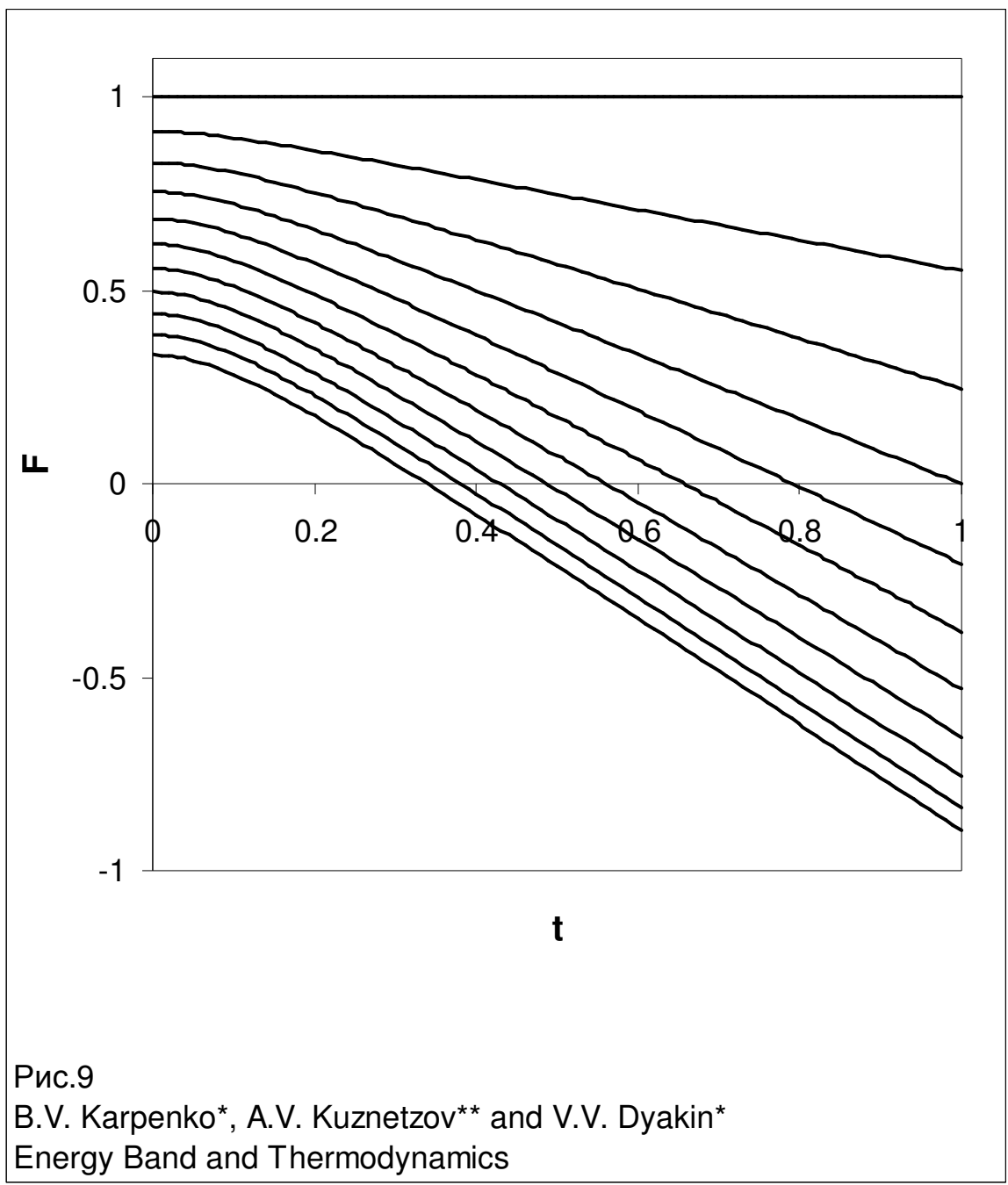

Рис.9
B.V. Karpenko*, A.V. Kuznetzov** and V.V. Dyakin*
Energy Band and Thermodynamics

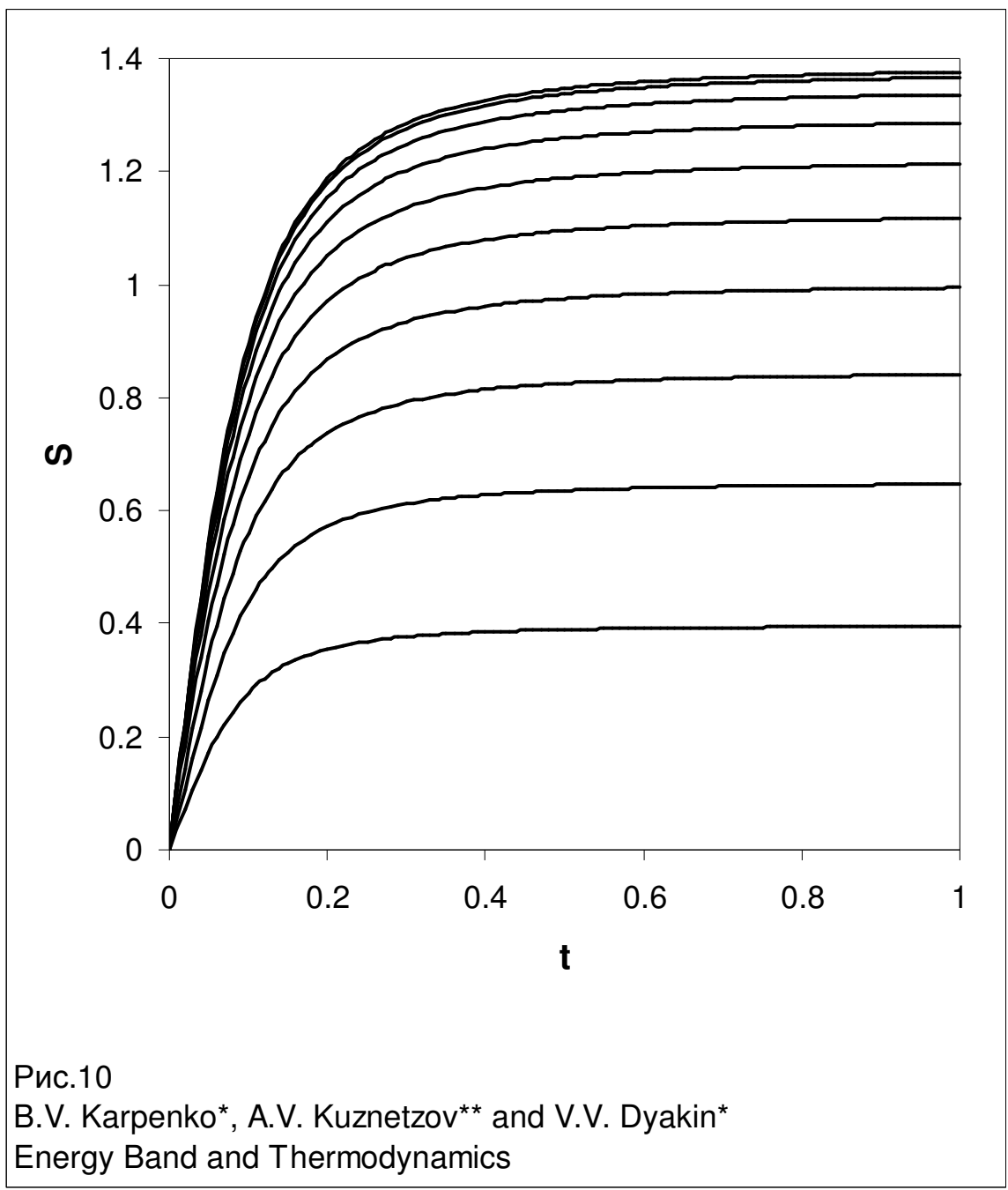

Рис.10
B.V. Karpenko*, A.V. Kuznetzov** and V.V. Dyakin*
Energy Band and Thermodynamics

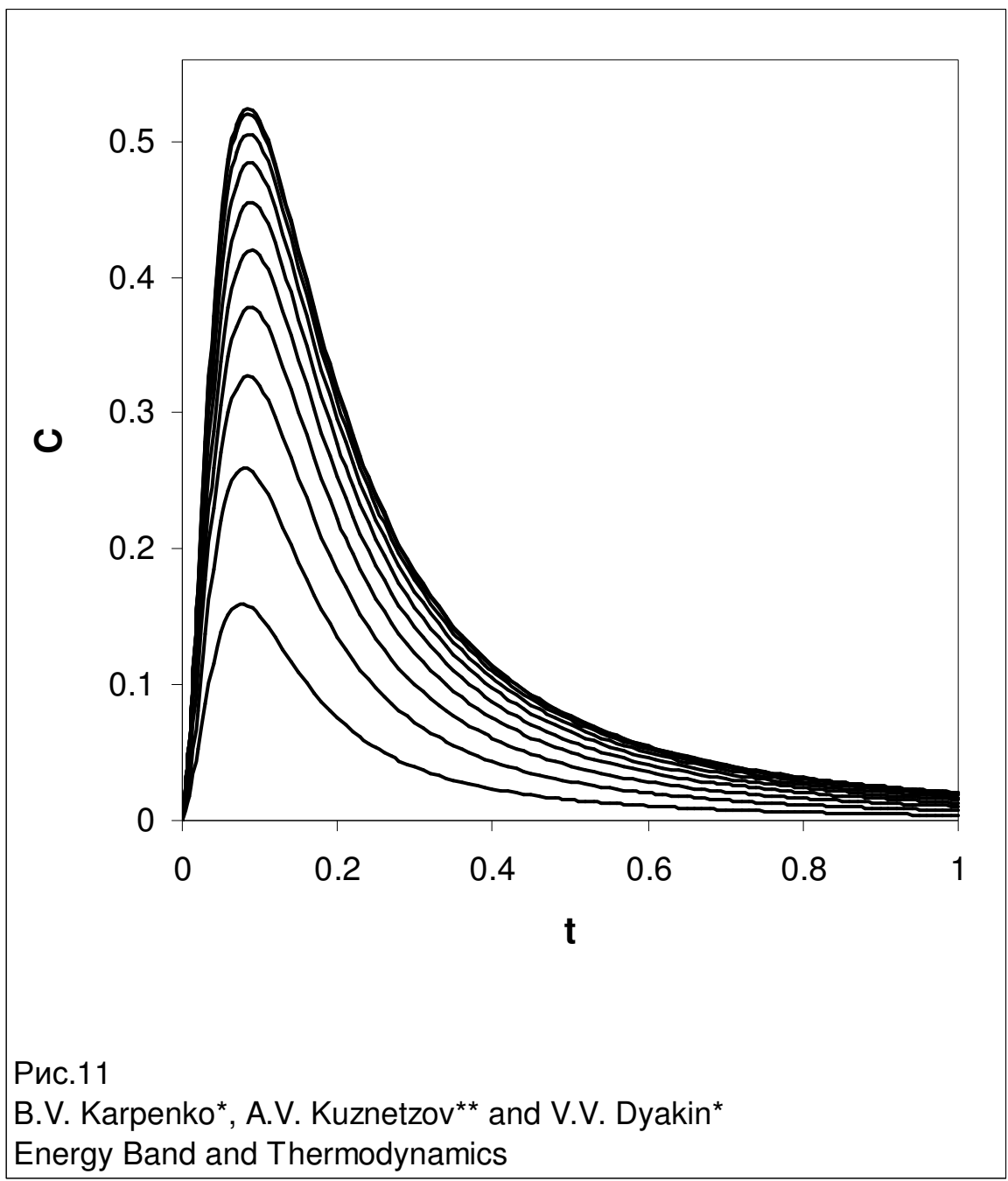

Рис.11
B.V. Karpenko*, A.V. Kuznetzov** and V.V. Dyakin*
Energy Band and Thermodynamics

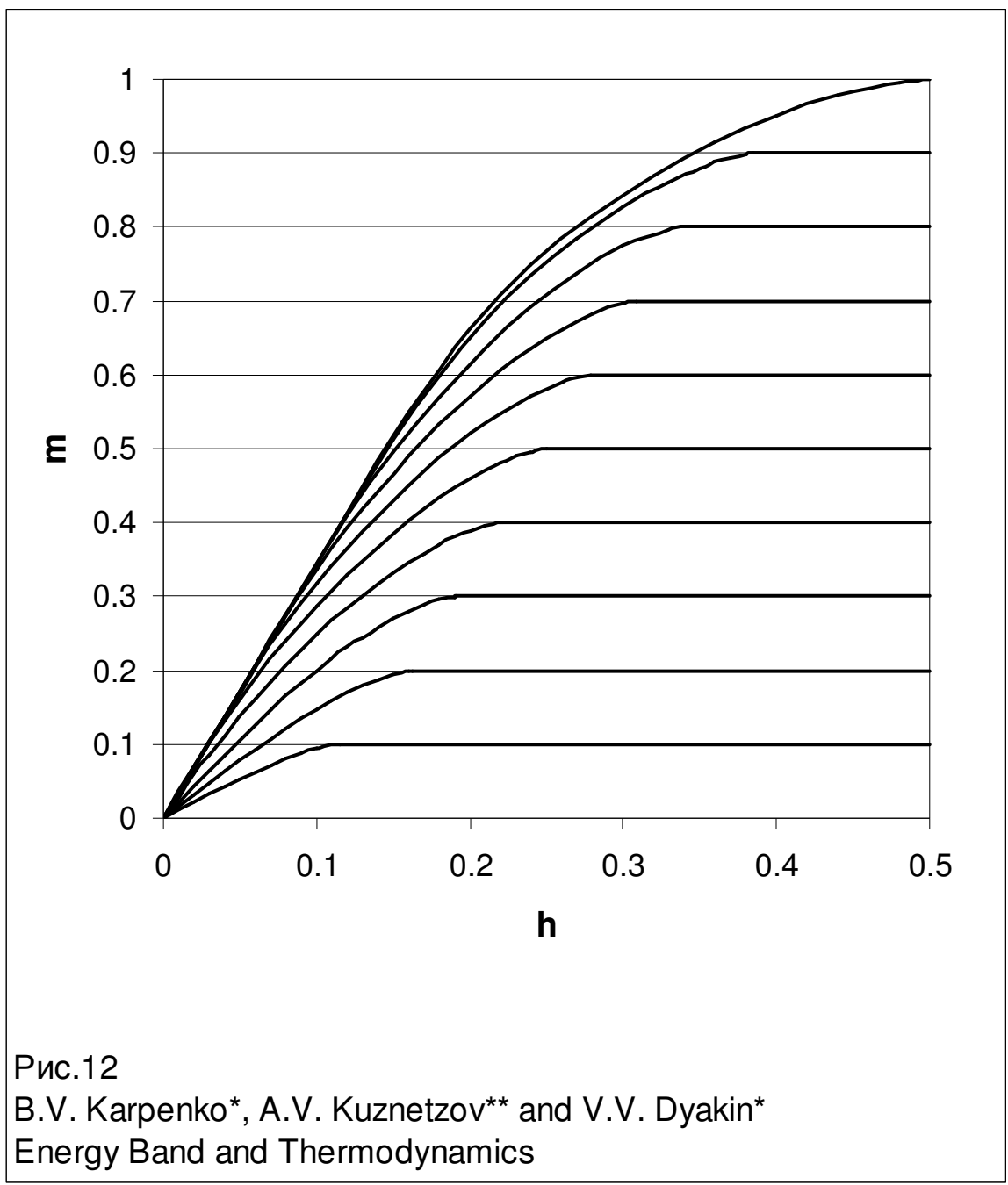

Рис.12
B.V. Karpenko*, A.V. Kuznetzov** and V.V. Dyakin*
Energy Band and Thermodynamics

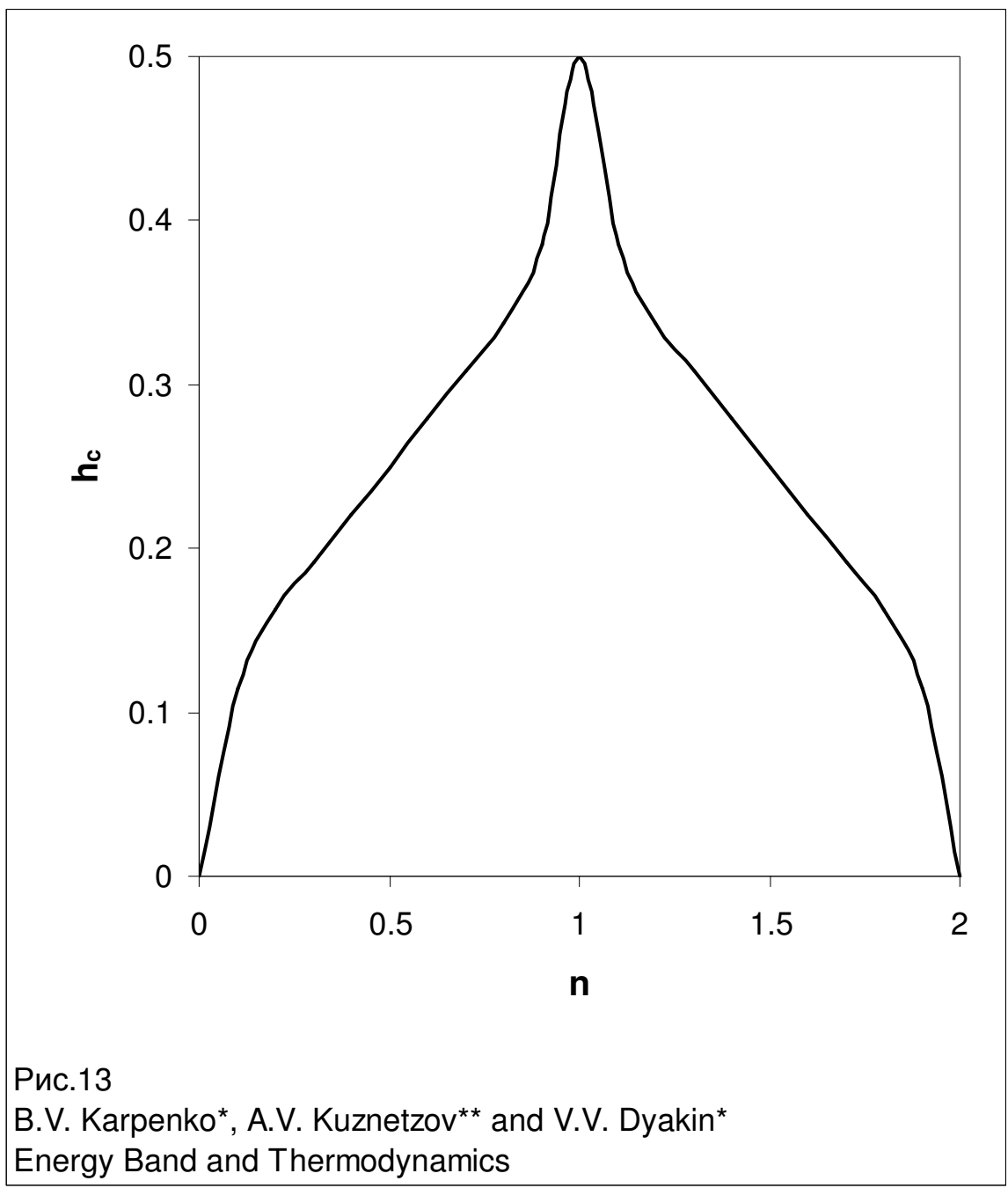

Рис.13
B.V. Karpenko*, A.V. Kuznetzov** and V.V. Dyakin*
Energy Band and Thermodynamics

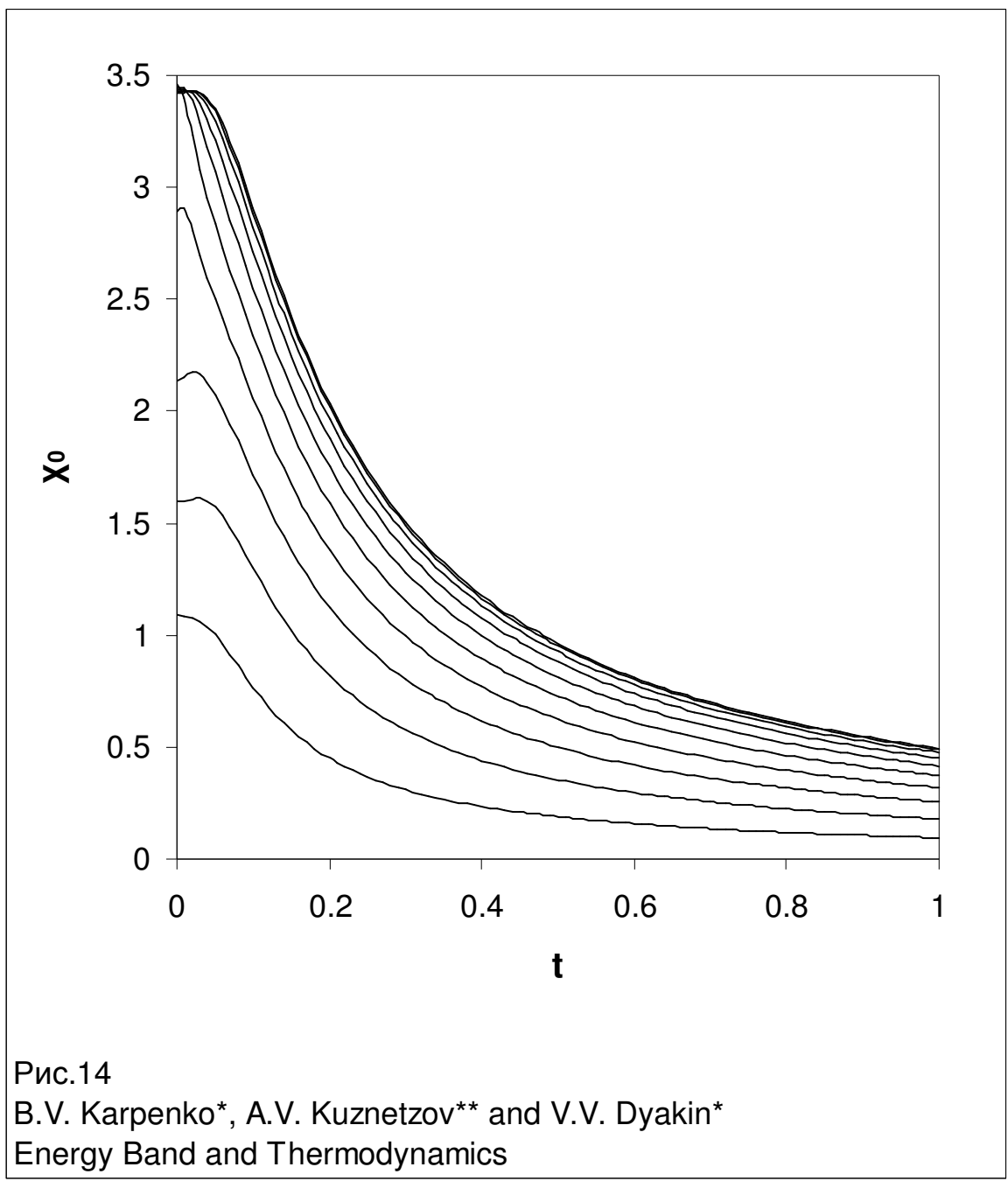

Рис.14
B.V. Karpenko*, A.V. Kuznetzov** and V.V. Dyakin*
Energy Band and Thermodynamics

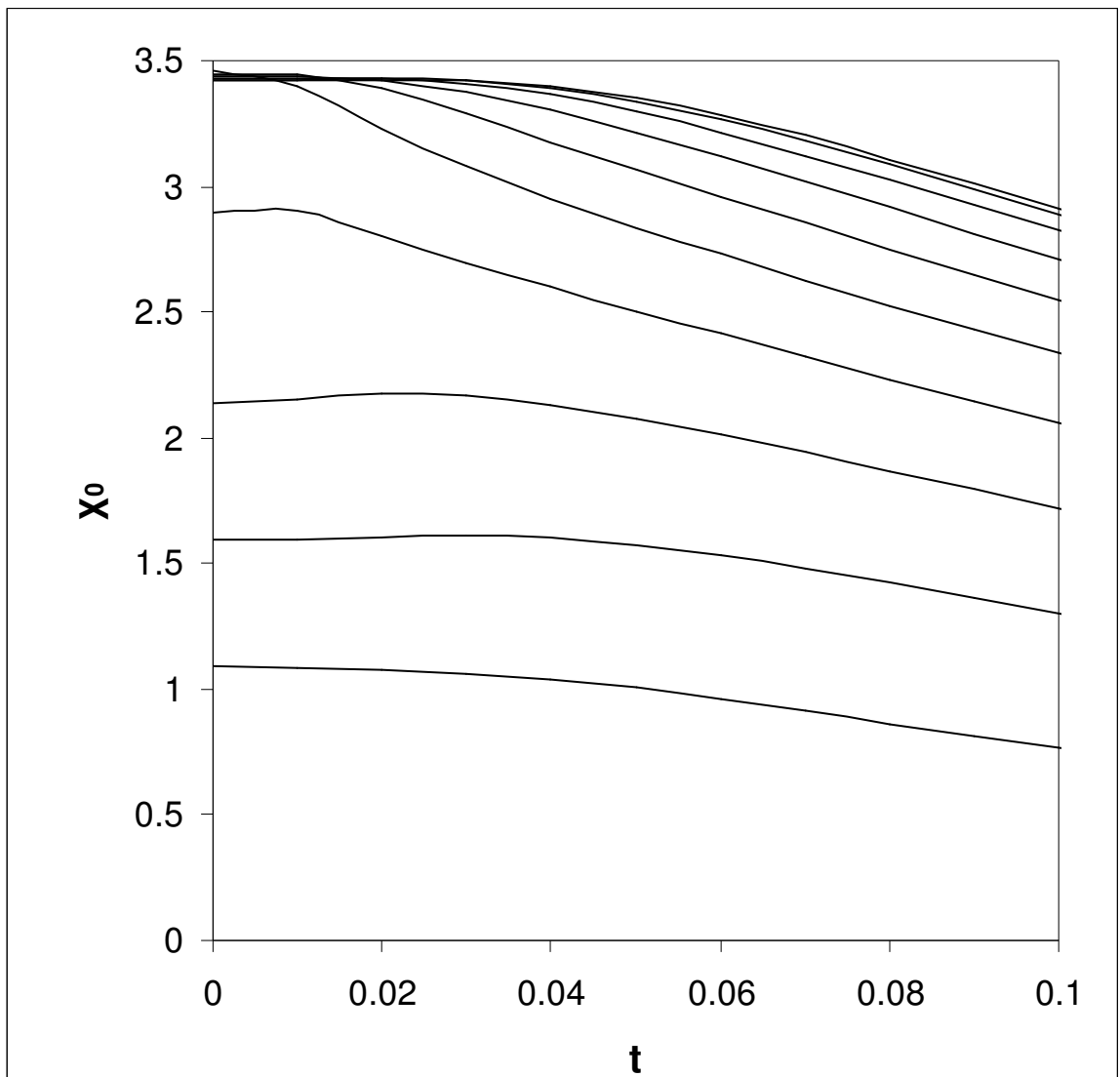

Рис.15
B.V. Karpenko*, A.V. Kuznetzov** and V.V. Dyakin*
Energy Band and Thermodynamics

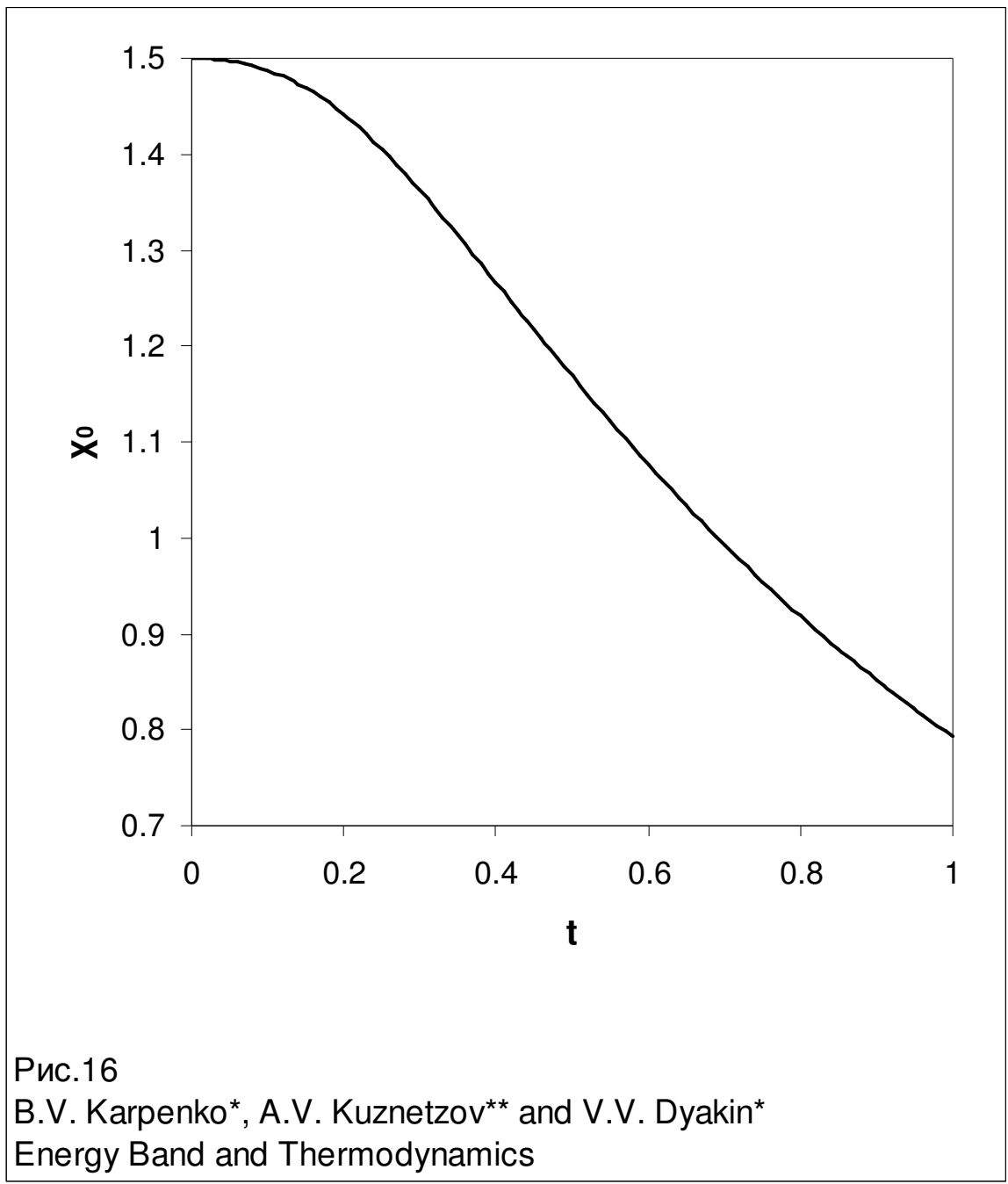

Рис.16
B.V. Karpenko*, A.V. Kuznetzov** and V.V. Dyakin*
Energy Band and Thermodynamics